\documentclass[11pt]{article}

 \usepackage{hyperref} \pdfoutput=1

\begin{document}

\title{Floating Extensional Flows}

\author{Roiy Sayag, Samuel S. Pegler, and M. Grae Worster \\ 
\\\vspace{6pt}
Department of Applied Mathematics and Theoretical Physics,\\ 
University of Cambridge, Cambridge, CB3 0WA, UK}

\maketitle
\begin{abstract}
  This fluid dynamics video demonstrate the breaking of axisymmetry in
  the floating extensional flow of a non-Newtonian fluid.
\end{abstract}

The centres of glacial ice sheets are grounded to the underlying bed and
dominated by shear flow. As the ice deforms under gravity it can
become sufficiently thin to float over surrounding oceans. These
floating regions of ice (ice shelves) are characterised by extensional
flows. The dynamical differences between the two regions can result in
substantial differences in the flow patterns, owing to the 
non-Newtonian ice rheology.

In this video
(\href{http://arxiv.org/src/1110.3283v2/anc/SPW-2011-DFD-Gallery.mov}{high}/\href{http://arxiv.org/src/1110.3283v2/anc/SPW-2011-DFD-Gallery_web.mov}{low}
resolution) we model the flow of an ice sheet into a denser ocean with
a viscous gravity current that intrudes into a denser salt
solution. To emphasise the important role of rheology in the floating
extensional region, we compare and contrast a Newtonian flow to a
non-Newtonian flow in two parts. First, we present a viscous gravity
current on a flat plane released from a point source at a constant
flux. Both the Newtonian and non-Newtonian fluids demonstrate an
axisymmetric front. In the Second part, the viscous gravity currents
intrude a denser ocean and a floating region forms. The Newtonian
fluid maintains axisymmetric flow in the floating region. In contrast,
axisymmetry breaks down in the non-Newtonian extensional region, which
ruptures into a set of finger-like protrusions. These observations may
relate to the mechanism underlying the break-up of large ice shelves.

 \end{document}